# DARK CURRENT AND RADIATION SHIELDING STUDIES FOR THE ILC MAIN LINAC[*][†]

N.V. Mokhov[#], I.L. Rakhno, N.A. Solyak, A. Sukhanov, I.S. Tropin

Fermi National Accelerator Laboratory, Batavia IL 60510-5011, USA

## ABSTRACT

*Electrons of dark current (DC), generated in high-gradient superconducting RF cavities (SRF) due to field emission, can be accelerated up to very high energies—19 GeV in the case of the International Linear Collider (ILC) main linac—before they are removed by focusing and steering magnets. Electromagnetic and hadron showers generated by such electrons can represent a significant radiation threat to the linac equipment and personnel. In our study, an operational scenario is analysed which is believed can be considered as the worst case scenario for the main linac regarding the DC contribution to the radiation environment in the main linac tunnel. A detailed modelling is performed for the DC electrons which are emitted from the surface of the SRF cavities and can be repeatedly accelerated in the high-gradient fields in many SRF cavities. Results of MARS15 Monte Carlo calculations, performed for the current main linac tunnel design, reveal that the prompt dose design level of 25 µSv/hr in the service tunnel can be provided by a 2.3-m thick concrete wall between the main and service tunnels.*

[*]Work supported by Fermi Research Alliance, LLC under contract No. DE-AC02-07CH11359 with the U.S. Department of Energy.

[†]Presented paper at the 13th Meeting of the task-force on Shielding aspects of Accelerators, Targets and Irradiation Facilities (SATIF-13), HZDR, October 10-12, 2016, Dresden, Germany

[#]mokhov@fnal.gov

# DARK CURRENT AND RADIATION SHIELDING STUDIES FOR THE ILC MAIN LINAC

**Nikolai V. Mokhov, Igor L. Rakhno, Nikolay A. Solyak, Alexander Sukhanov, Igor S. Tropin**
Fermi National Accelerator Laboratory, Batavia, Illinois 60510, USA

## Introduction

During design and optimization of an SRF linac, one has to inevitably deal with dark current (DC) issues. Dark current electrons, generated in a cryo-module (CM), produce radiation that affects various elements nearby: (i) beam line components and cables inside the CM; (ii) electronics outside of the CM in the linac tunnel; (iii) electronics and personnel in the service part of the linac tunnel. Therefore, extensive studies of DC radiation are required during the design of the SRF linac in order to protect various accelerator components from radiation damage and optimize thickness and cost of radiation shielding. Similar DC issues were addressed at Fermilab about a decade ago when a vertical test stand for SRF cavities was under construction [1,2]. A distinctive feature of those studies was limited number of SRF cavities—at any given time the test facility could contain no more than six such cavities. From physical standpoint, the DC generation mode is the same in both the previous and current studies—RF field-induced quantum tunnelling from the SRF cavity surfaces initially studied and described by Fowler and Nordheim [3]. From a technical standpoint, however, the current study for the ILC main linac is more complicated because it deals with sections of the linac consisting of up to 40 basic RF units (periods), and every single unit contains 4 cryo-modules, 26 SRF cavities and a focusing quadrupole magnet. A detailed description of the generated distribution of the DC electrons, that serves as an input for subsequent Monte Carlo modelling of interactions with materials of accelerator components, is given in Ref. [4]. The modelling of interactions with matter is performed with the MARS15 code [5,6]. In order to normalize the generated DC and all subsequent calculations, one assumes that in every single SRF cavity the current is equal to 50 nA. The study was performed with TESLA-type 9-cell 1.3-GHz SRF cavities.

## Field emission and particle tracking in SRF cavities

Experimental observations of field-induced emission (in other words—dark current) in SRF cavities can be summarized as follows: (i) the field-induced emission is generally the result of various imperfections, *e.g.*, residual dust contamination in the cavity; (ii) the imperfections can give rise to a significant enhancement of local electric field and, consequently, field-emitted electrons which can generate secondary gammas in surrounding materials; (iii) the surface imperfections can happen anywhere, but field emission occurs mostly around irises–locations with the highest local electric field; (iv) for a given SRF cavity the emission usually does not occur at several sites; it usually happens at a single site and lasts until a significant amount of RF energy stored in the cavity is lost to the generated DC.

In our model, the field-emitted electrons are modeled inside the cavity until they hit the cavity surface. The DC generation is assumed to be uniform over the cavity surface, and a weight technique is employed. For a given region, the probability of field-induced emission depends greatly on the magnitude and RF phase of the surface electric field [7]. Therefore, the relative probabilities of possible electron trajectories differ significantly, and the most probable trajectories usually do not correspond to the highest electron energy gain in the accelerating field. The field emission is generated with azimuthal symmetry. Phase-space coordinates of such events represent a digitized source term for subsequent modeling of secondary particle generation and transport in the entire system with the MARS15 code. A detailed description of the generated distribution of the DC electrons, that serves as an input for subsequent Monte Carlo modelling, is given in Ref. [4]. Various modelled DC electron trajectories in the SRF cavities are shown in Figure 1.

A detailed modeling is performed for the DC electrons which are emitted from the surface of the SRF cavities and can be repeatedly accelerated in the high-gradient fields in many cavities. Results of our modelling in the SRF cavity fields can be summarized as follows:

- About 92% of emitted particles are absorbed in the same cavity where they were generated ;
- 4% of emitted particles are captured into an acceleration mode and exit the cavity in the direction of the main beam thus entering the next cavity ;
- The remaining 4% of emitted particles exit the cavity upstream, that is in the direction opposite to the main beam thus entering the previous cavity.

**Figure 1: Various modelled DC electron trajectories**

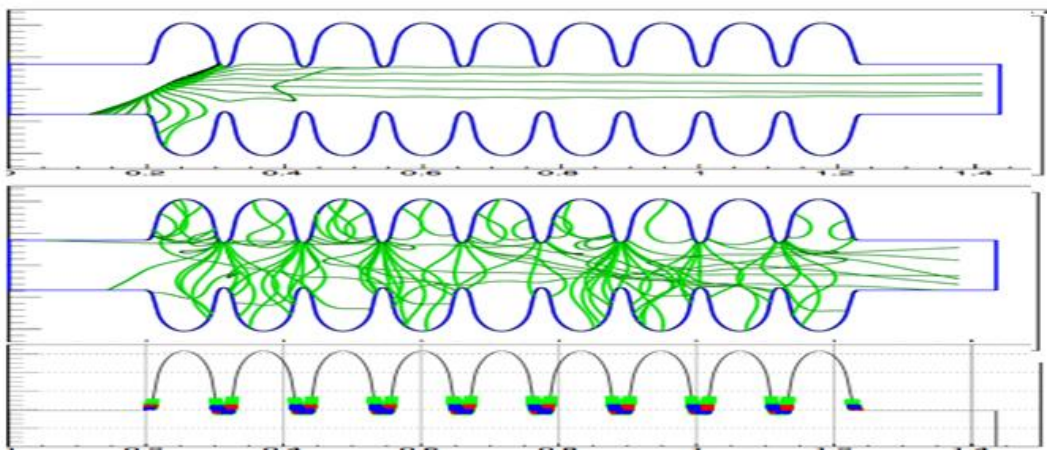

**MARS15 geometry model**

A cross section of the ILC Kamaboko tunnel is shown in Figure 2. In the current design, there is a wall between main and service tunnels. Thickness of this wall (1.5-3.5 m), that separates the service and operational parts of the tunnel, is determined by the maximum beam losses. A reduction of the wall thickness is considered as a cost reducing option.

**Figure 2: A cross section of ILC Kamaboko tunnel (the dimensions are given in mm)**

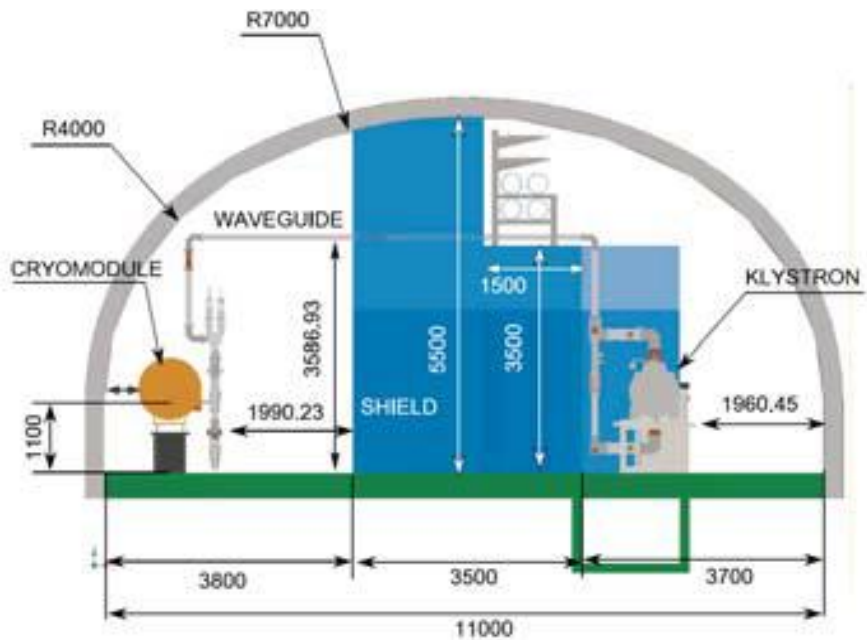

An advanced MAD-MARS beam line builder, that generates a 3D ROOT geometry [8] for linac sections under consideration, is used in these studies. The following energy thresholds were used: $10^{-3}$ eV for neutrons and 0.1 MeV for all other particles. Several fragments of the MARS15 geometry model are shown in Figure 3. Three-dimensional DC distributions, generated at the first stage of the study, served as an input for subsequent Monte Carlo modeling with the MARS15 code. One can see from the Figure that the current geometry model provides a quite realistic description of the ILC main linac with major beam line components and tunnel.

**Figure 3: Fragments of the MARS15 model – tunnel cross section (top left), SRF cavity (top right), cryo-module elevation view with an SRF cavity and cross section with a quadrupole magnet (bottom left and right, respectively)**

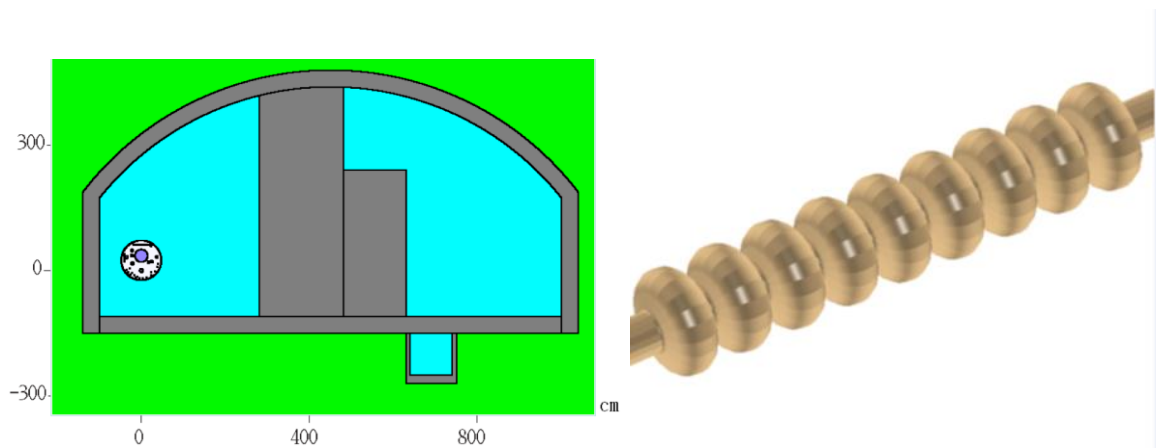

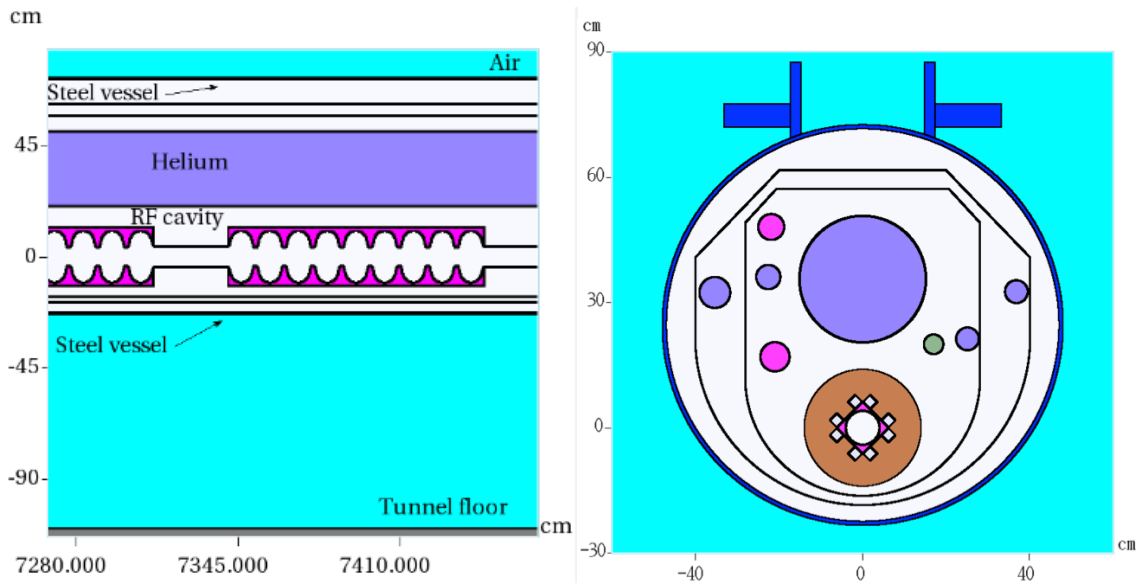

## Modes of operation

First, we considered a normal mode of operation of the linac, when both RF power to the cavities and quadrupole magnets were turned on. Also studied were a few commissioning modes of operation when only RF power to the cavities is turned on, while quadrupole magnets were turned off. It turned out that a commissioning mode can be more severe from the standpoint of prompt radiation. Therefore, the problem of selecting the worst case scenario actually boils down to the question of how realistic the assumption of equal 50 nA DC contribution from every single SRF cavity is.

### Normal mode

In the normal mode of operation, the DC electrons are lost in the magnets or in the cavity downstream. The maximum energy of the lost DC electrons can be as high as 800 MeV. In this case, the equilibrium state, when losses of DC particles along the linac are compensated by newly generated DC electrons, is reached already at the second RF unit. The largest losses in this mode are observed at the end of the linac and concentrated at the locations of focusing magnets. Figure 4 shows the distribution of the prompt dose over the tunnel cross section at the focusing magnet location at the end of linac. The dose is as low as 25 µSv/hr after a 1.2-m thick concrete wall. Thus, in the current design of the ILC main linac and in the normal mode of operation, the 3.5-m concrete wall between main and service tunnels provides a very large safety margin for radiation protection.

**Figure 4: Prompt dose (mSv/hr) over the tunnel cross section at the end of the ILC main linac for normal operation**

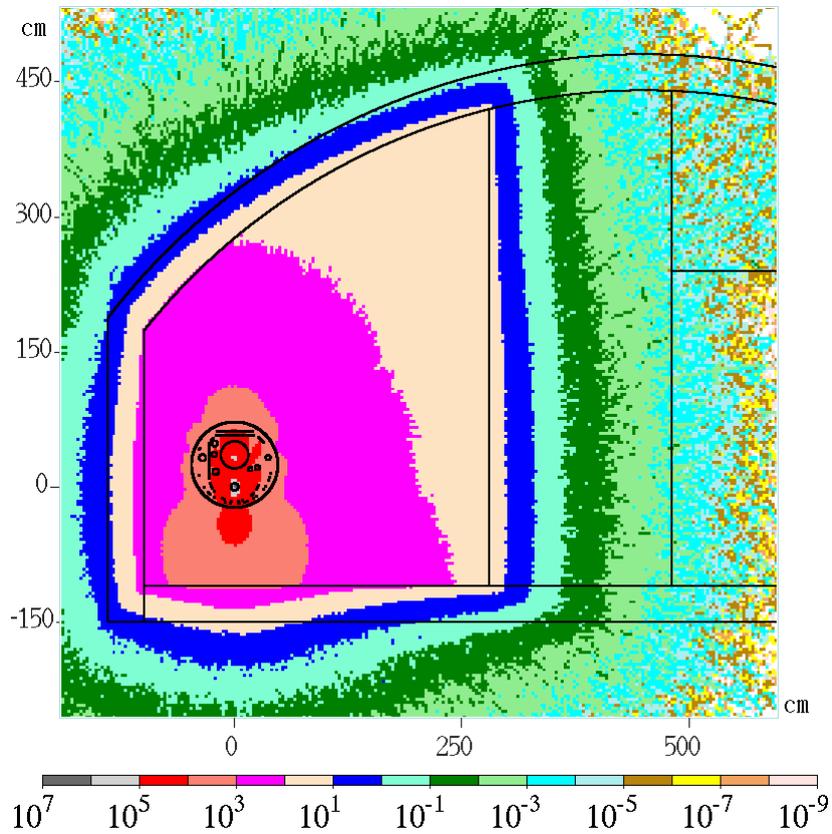

*Commissioning modes*

Turning off quadrupole magnets while still having RF power in cavities can, potentially, present worse conditions from standpoint of radiation generated around the linac when comparing to the normal mode of operation described above. We considered four different commissioning modes:

- A straight section of the linac (bunch compressor) with steering/correcting magnets turned off.

- A curved section of the linac, which follows Earth curvature, with steering/correcting magnets turned off.

- A curved section with perfect alignment, steering magnets on, but no correctors off.

- A curved linac with random misalignment, with steering magnets and correcting magnets on.

When quads are turned off, DC electrons can traverse many RF units of linac before they hit a material and are absorbed. Therefore, an equilibrium state—from the standpoint of losses—can be reached only after multiple RF units, and energies of these DC electrons after multiple accelerations can be pretty high. In our study, we track DS particles through 40 RF units (1.5 km) of the linac. Results of our calculations reveal that the **worst case** is the last option listed—**a curved linac with** **steering magnets on and correction on for misalignment**, see Figure 5. The loss in the plateau region—beyond 800 m—was used to build the source for MARS15 modelling. One can see that it takes about 700 m to reach a uniform beam loss distribution. Calculated energy spectra of the DC electrons for the worst case are shown in Figure 6. Figure 7 gives the total prompt dose for the worst case, and Figure 8 provides partial neutron and photon contributions to the total dose. In Figure 9 one compares the normal operation mode and the worst case during commissioning. One can see that in the worst case the radiation level is approximately an order of magnitude higher, and that the dose level of 25 µSv/hr is observed at the depth of 2.2 m in the concrete wall. For the worst-case commissioning mode, Figure 10 shows 2D maps of prompt dose (mSv/hr) and yearly absorbed dose (Gy/yr) across a cryo-module in the plateau region.

**Figure 5: Power loss per cryo-module along the curved ILC linac with steering magnets turned on and correctors off (red) or on (blue). Every single SRF cavity contributes 50 nA dark current**

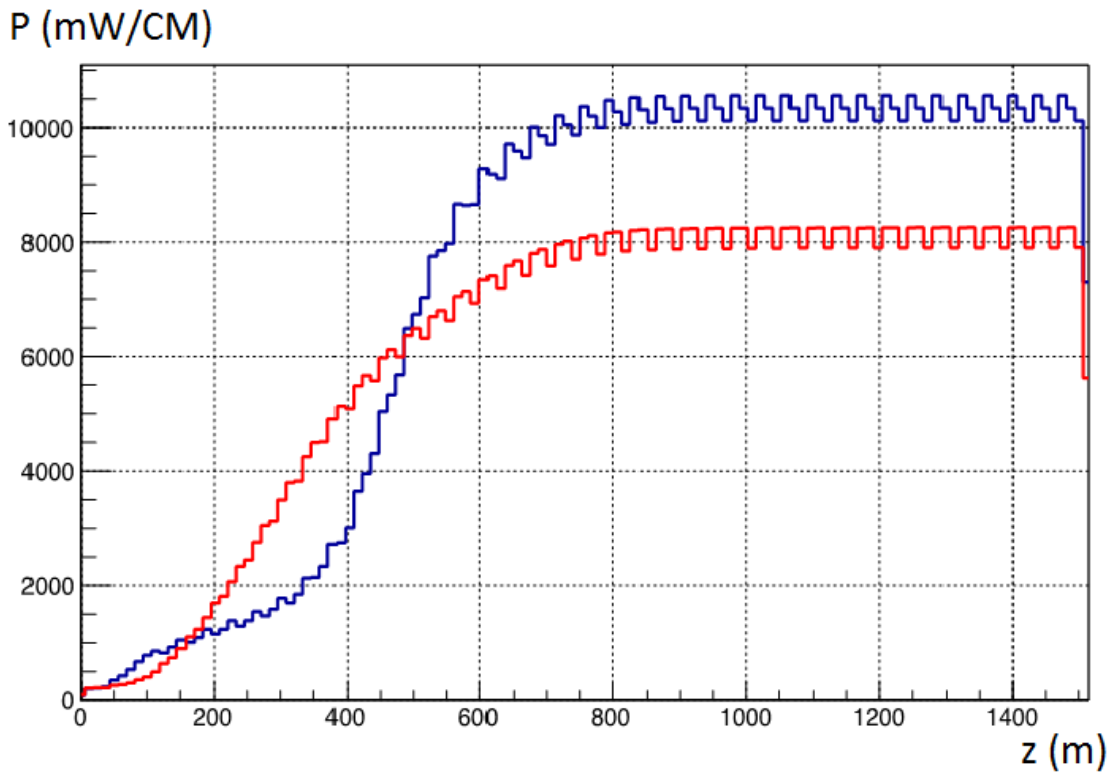

Figure 6: Electron energy spectra for high-energy (top) and low-energy (bottom) parts in the file used for MARS15 modelling as a source corresponding to the worst case during commissioning. Average energy is equal to 2.4 GeV and 1.5 MeV for the high-energy and low-energy parts, respectively. The highest energy observed in the spectra is equal to 19.2 GeV and 28 MeV, respectively

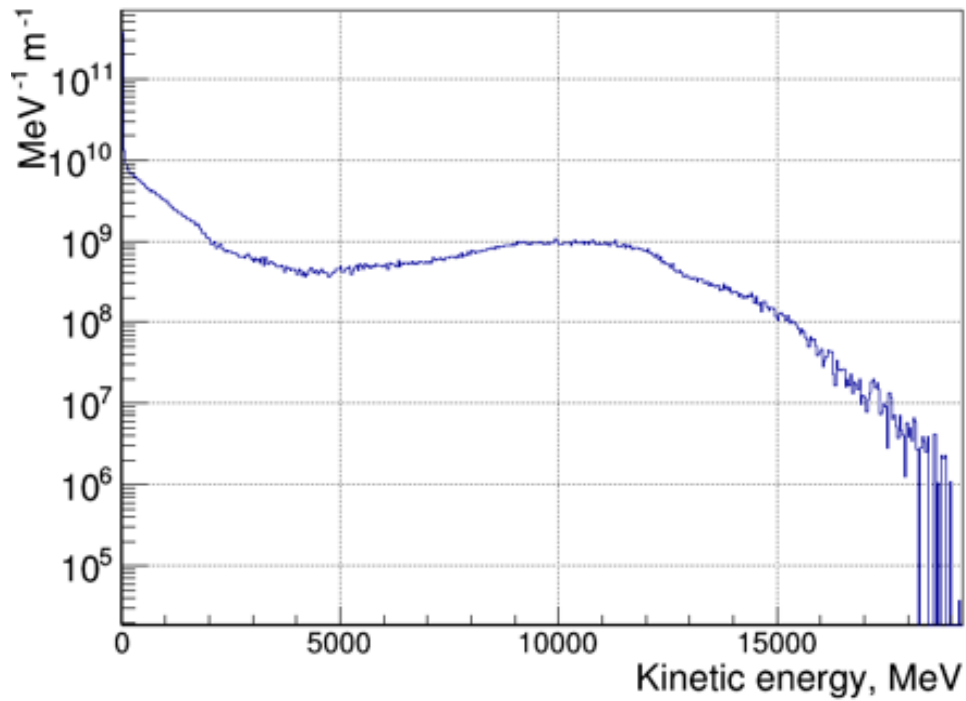

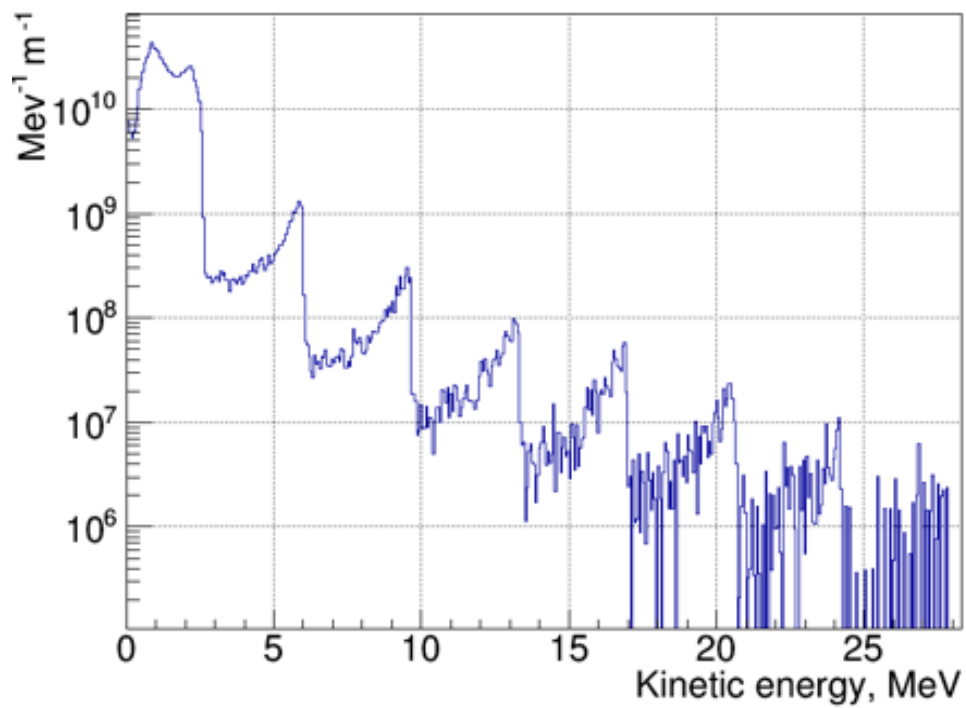

**Figure 7: Total prompt dose (mSv/hr) over the tunnel cross section at the end of the ILC main linac for the worst case during commissioning**

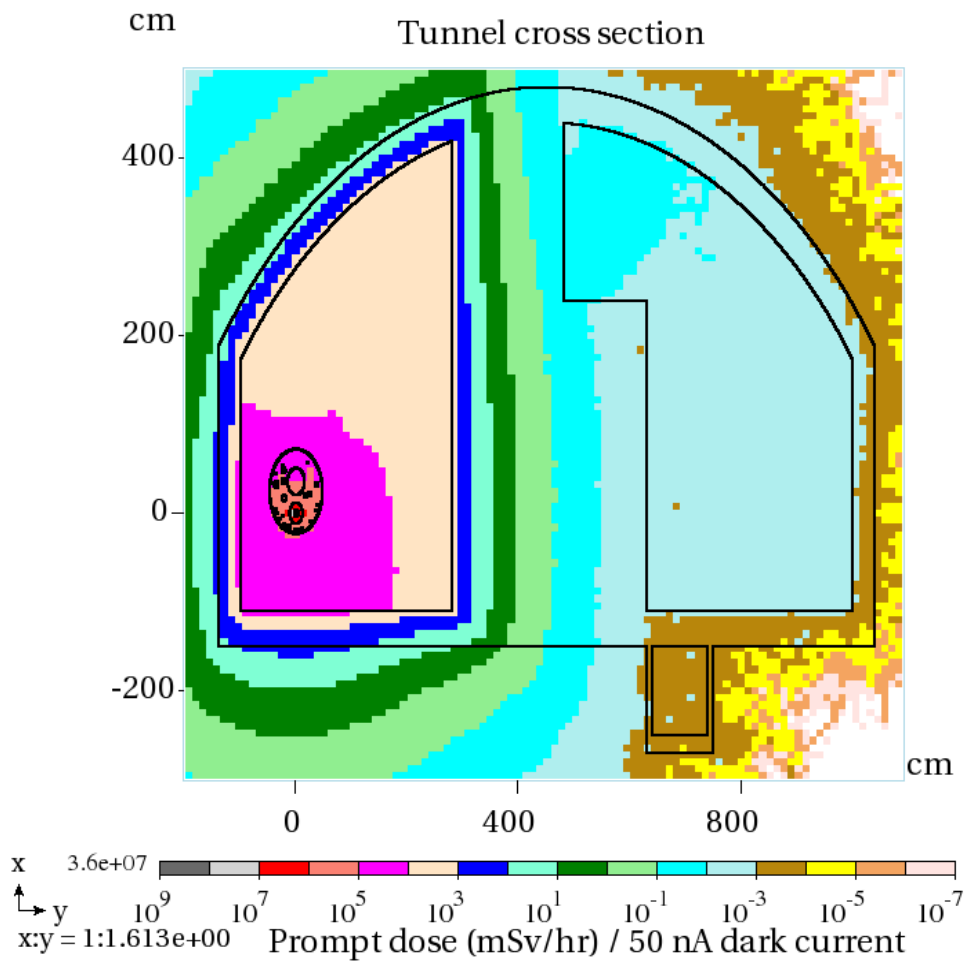

**Figure 8: Partial photon (left) and neutron (right) prompt dose (mSv/hr) over the tunnel cross section at the end of the ILC main linac for the worst case during commissioning**

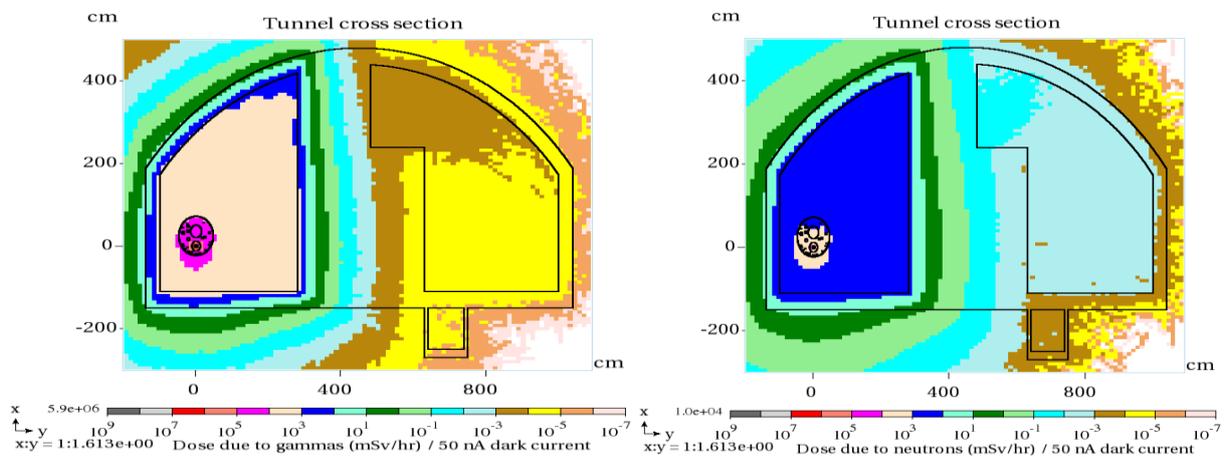

Figure 9: A comparison between normal operation mode (case A) and the worst case during commissioning (case B) is shown in upper part. In both the cases the quads correspond to a 250-GeV beam. Lower part shows locations of the maximum prompt dose in the tunnel observed in both the cases

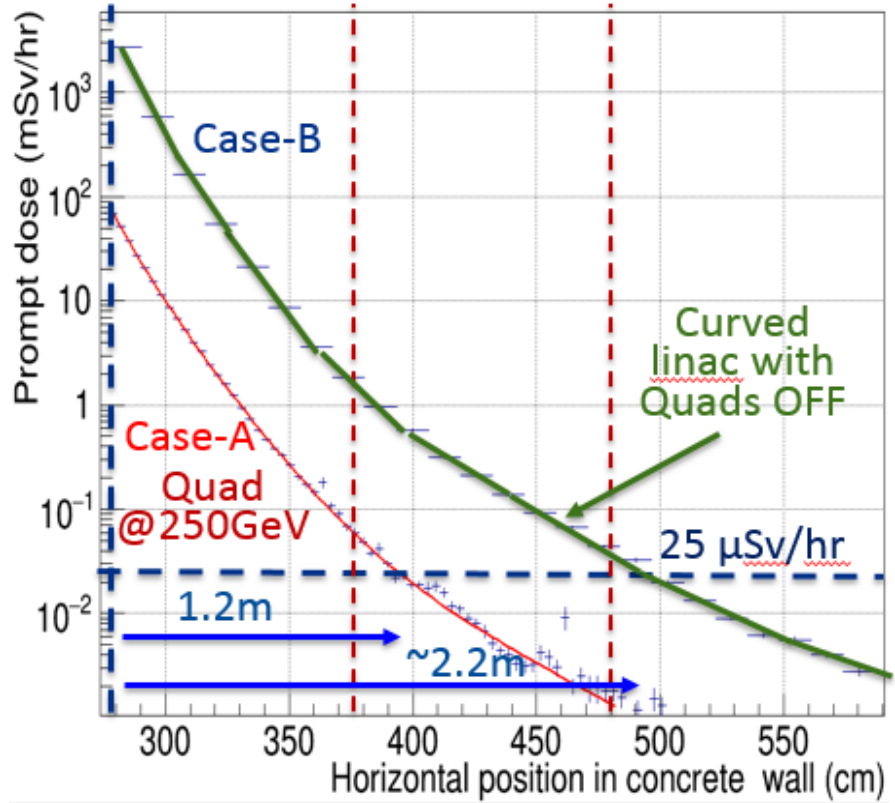

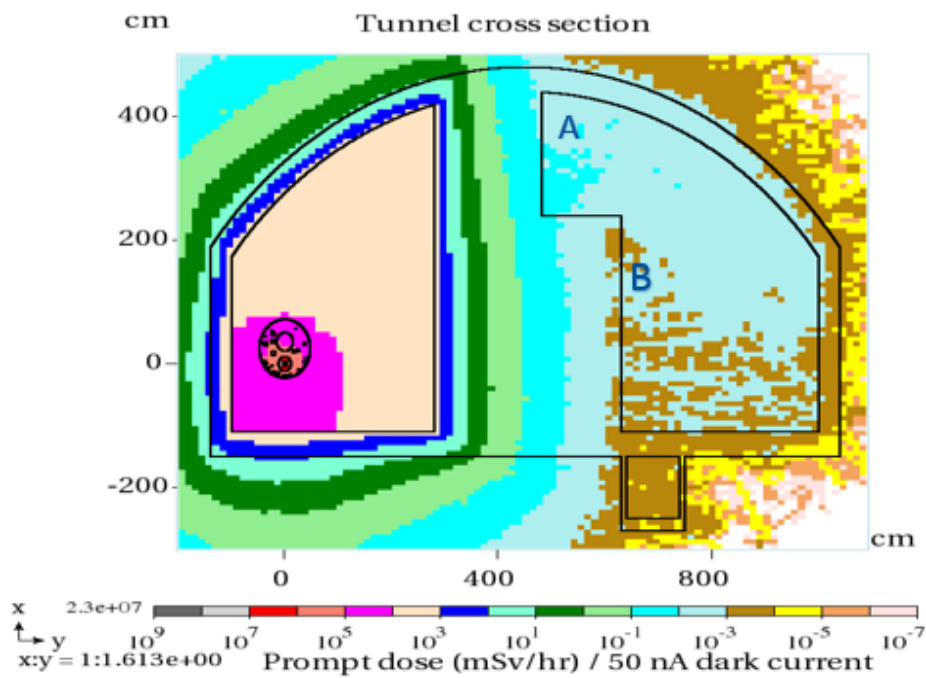

Figure 10: Total prompt (top) and absorbed (bottom) dose calculated for a cryo-module for the worst-case commissioning mode.

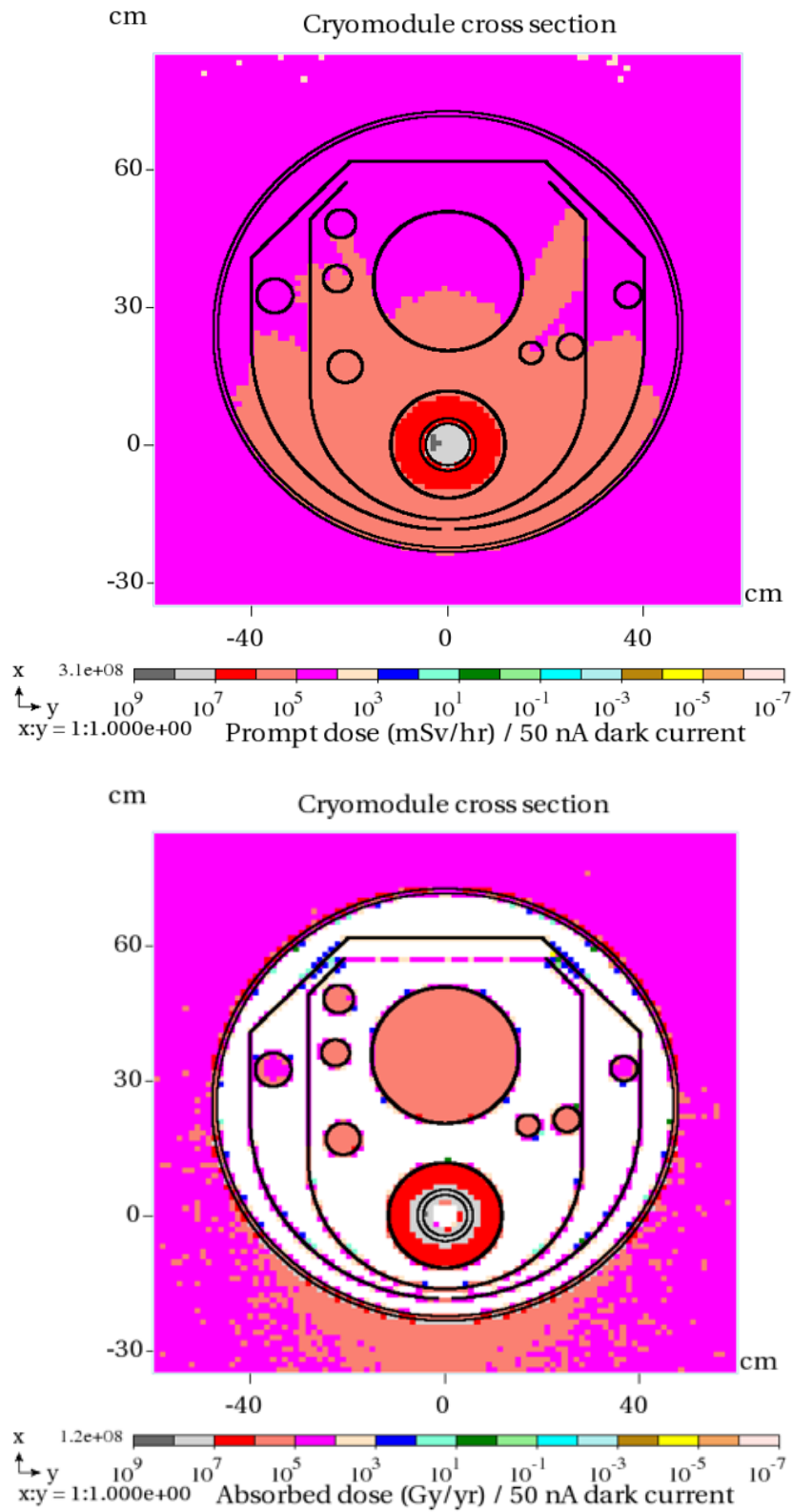

## Conclusions

An elaborate model of dark current, generated in SRF cavities, is employed to study DC radiation in the ILC main linac tunnel. A normal operation and commissioning mode of the linac are considered, when focusing magnets are turned on and off, respectively. In order to normalize the generated DC, we assume that every single SRF cavity contributes 50 nA of DC. Tracking of the generated DC electrons is performed in the linac sections as long as 1.5 km (40 basic RF periods). Results of this tracking are used as a source for subsequent MARS15 modelling of interaction of the radiation with matter to calculate radiation levels in the tunnel.

Results of our calculations reveal that, for approximately half of the entire linac tunnel, the radiation environment is practically independent of the longitudinal position. The major contribution to prompt dose behind the tunnel wall is due to secondary neutrons. The worst case scenario is a commissioning mode with focusing magnets turned off and both steering and correcting magnets are on. In this case, the prompt dose at the surface of the wall in the main tunnel is approximately 2000 mSv/hr, and the dose drops to the design level of 25 µSv/hr after 2.2 m of the concrete wall. Therefore, the current design of the ILC main linac with concrete wall of 3.5 m provides a large safety margin to protect the personnel and electronics in the service tunnel.

## Acknowledgements

This work is supported by Fermi Research Alliance, LLC under contract DE-AC02-07CH11359 with the U.S. Department of Energy.